\documentclass[prl,twocolumn,showpacs]{revtex4}
\usepackage{graphicx}
\usepackage{amsmath}
\usepackage{amsfonts}
\usepackage{amssymb}
\usepackage{epsfig}

\begin{document}

\title{Non-invasive vibrational mode spectroscopy of ion Coulomb crystals through resonant collective coupling to an optical cavity field}

\author{A. Dantan, J. P. Marler, M. Albert, D. Gu\'{e}not}
\affiliation{QUANTOP, Danish National Research Foundation Center for
Quantum Optics, Department of Physics and Astronomy, University of
Aarhus, DK-8000 \AA rhus C., Denmark}

\author{M. Drewsen}\email{drewsen@phys.au.dk}
\affiliation{QUANTOP, Danish National Research Foundation Center for
Quantum Optics, Department of Physics and Astronomy, University of
Aarhus, DK-8000 \AA rhus C., Denmark}

\begin{abstract}
We report on a novel non-invasive method to determine the normal
mode frequencies of ion Coulomb crystals in traps based on the
resonance enhanced collective coupling between the electronic states
of the ions and an optical cavity field at the single photon level.
Excitations of the normal modes are observed through a Doppler
broadening of the resonance. An excellent agreement with the
predictions of a zero-temperature uniformly charged liquid plasma
model is found. The technique opens up for investigations of the
heating and damping of cold plasma modes, as well as the coupling
between them.
\end{abstract}


\date{\today}

\maketitle

Cold one-component plasma physics~\cite{dubin99} has in the past two
decades led to a series of interesting results due to the
availability of fast
computers~\cite{schiffer96,schiffer00,totsuji02,matthey03} as well
as the possibility to experiment with ensembles of trapped,
laser-cooled atomic
ions~\cite{itano98,bollinger98,wineland87,walther92,drewsen98,mortensen06,mortensen07,heinzen91,bollinger93,mitchell98,jensen05}.
Prominent examples are the understanding of structural properties of
crystallized cold plasmas in both Penning~\cite{itano98,bollinger98}
and
Paul~\cite{wineland87,walther92,drewsen98,mortensen06,mortensen07}
traps, and the investigation of the normal mode dynamics of cold
magnetized plasmas in Penning
traps~\cite{heinzen91,bollinger93,mitchell98,jensen05}. While there
exist many similarities between experiments in Penning and Paul
traps, the unmagnetized plasmas in Paul traps are e.g. known to heat
up much faster than the magnetized plasmas in Penning traps due to
the presence of the rf fields \cite{blumel89,schiffer00}. The lack
of a rotational symmetry axis in Paul traps has as well been found
to be responsible for the observation of specific crystalline
structures~\cite{mortensen07}. Exploration of the normal mode
dynamics of unmagnetized ion Coulomb crystals in e.g. linear Paul
traps will hence add to our understanding of the influence of the
trapping environment on the physics of such crystals. Furthermore,
since large Coulomb crystals are excellent candidates for the
realization of high-fidelity quantum memories for
light~\cite{herskind09}, such studies can shed light on the
influence of the excitation of these modes on the fidelity as well
as on the prospect of storing several photonic quantum bits through
deliberate excitations of specific vibrational modes. Larger Coulomb
crystals have as well recently been considered as a system for
performing quantum simulations, in which respect knowledge of normal
mode dynamics is needed~\cite{porras04}. Finally, ion Coulomb
crystals represent extremely interesting systems to study cavity
optomechanics phenomena~\cite{murch08,brenneke08}, since, in spite
of their solid nature, they possess free atomic resonance properties
and can hence be made very sensitive to the radiation pressure force
exerted by optical fields.

In this Letter, we report on the study of normal mode vibrations of
Coulomb crystals of $^{40}$Ca$^+$ ions in a linear Paul trap by a
novel non-invasive technique. The technique is based on monitoring
the coherent collective resonant response of the atomic ions
constituting the crystal to a single photon optical cavity
field~\cite{herskind09}. By having a standing wave optical cavity
incorporated in the trap setup with the light propagation axis
coinciding with the rf-field free axis of the trap, we can detect
very small changes in the optical response due to Doppler shifts
resulting from the ions' motion along the cavity axis when a mode is
excited. Since the probing does not rely on the observation of
incoherently scattered photons as e.g. the case in Doppler
velocimetry~\cite{heinzen91,bollinger93,mitchell98} and in
Sympathetically-Cooled Single Ion Mass
Spectrometry~\cite{drewsen04}, the detected signal can in principle
be purely dispersive in nature and does not require any excitation
of the ions. The technique is not limited to linear Paul traps, but
should work as well in any open structured trap geometries such as
e.g. the Penning-Malmberg type traps \cite{driscoll88,heinzen91}.

Previous Doppler velocimetry imaging experiments with cold
magnetized plasmas of Be$^+$ ions in a Penning trap have led to the
identification of a series of normal
modes~\cite{heinzen91,bollinger93,mitchell98}, which correlate to
specific $(l,m)$-modes theoretically predicted for spheroidal
shaped uniformly charged liquids~\cite{dubin91}. For the cold
unmagnetized plasmas in the linear Paul trap used in our
investigations, the corresponding low-order normal mode frequencies
are expected to be close to those of the $(l,m)$ modes, which are
given by~\cite{dubin91}
\begin{equation}\label{eq:modefrequencies} \omega_{(l,m)}=\frac{\omega_p}{\sqrt{1-\frac{P_l^m Q_l^{m '}}{Q_l^m P_l^{m '}}}},\end{equation} where
$\omega_p=\sqrt{q^2n_0/M\varepsilon_0}$ is the plasma frequency (q
is the charge of the ion, M its mass and $n_0$ the uniform ion
density), $P_l^m=P_l^m(1/\sqrt{1-\alpha^2})$ and
$Q_l^m=Q_l^m(1/\sqrt{1-\alpha^2})$ are first and second-order
Legendre polynomials with cylindrical indices $(l,m)$ and with the
prime denoting differentiation with respect to the entire argument.
Finally, $\alpha$ is the aspect ratio of the plasma given by
$\alpha=R/L$, where $L$ and $R$ are the half-length and radius of
the spheroid, respectively (see also Fig.~\ref{fig2}a).

The corresponding spatial structure of the modes can generally be
expressed in terms of generalized spheroidal coordinates
~\cite{dubin96}. While the modes generally have a non-trivial
dependence on the ions' position in the crystal, for longitudinal
modes ($l,m=0$), the displacement $\delta z$, close ($r\simeq 0$) to
the axis of rotational symmetry ($z$-axis), is approximately given
by
\begin{equation}\label{eq:deltaz} \delta z\varpropto
P_l^{0'}(z/\sqrt{L^2-R^2}).\end{equation} Consequently, all modes
with $m=0$ have a spatial variation along the $z$-axis which, under
excitation, will lead to measurable Doppler-shifts of the ions
resonance frequency along this axis.
\begin{figure}
  \includegraphics[width=7cm]{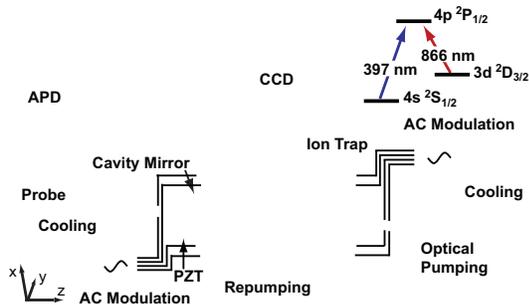}
  \caption{(color online.) Schematic of the linear Paul trap with an integrated optical cavity.
  The excitation of the crystal modes (AC modulation) can be monitored either by measuring the change in the reflection spectrum
   of a weak cavity field (866 nm) with an avalanche photodiode (APD) or by imaging the fluorescent
  light (397 nm) emitted by the $^{40}$Ca$^+$ ions during laser cooling onto a CCD camera.
  The inset shows the relevant energy levels and transitions of the $^{40}$Ca$^+$ ion.} \label{fig1}
\end{figure}

Figure~\ref{fig1} shows a schematic of the experimental setup for
our studies (a more detailed description of the cavity-trap setup
can be found in Ref.~\cite{herskind08}). The linear Paul trap
consists of four cylindrical electrodes to which an rf voltage
$U_{rf}$ between 200 V and 300 V is applied at a frequency $\Omega$
of 2$\pi\times$4 MHz for radial confinement ($xy$-plane in
Fig.~\ref{fig1}). To obtain a linear quadrupole configuration, the
rf voltages are $180^{\circ}$ out of phase between nearest neighbor
electrodes. Each of the four electrodes are sectioned into three
parts to allow for the application of DC (1-10 V) and AC voltages to
the eight end-pieces for static confinement of the ions along the
cavity axis ($z$-axis) and for normal mode excitations,
respectively. The electrodes have diameters of $5.2$ mm and the
nearest neighbour electrode spacing is $1.8$ mm. With these
parameters, the single ion effective (pseudo) potential is to a very
good approximation harmonic in all dimensions in the region of the
crystals, which leads to constant densities of the crystallized
$^{40}$Ca$^+$ ions with values between $2.6\times 10^8$ and
$5.7\times 10^8$ cm$^{-3}$~\cite{herskind09b}.

The $^{40}$Ca$^+$ ions are produced through isotope selective
photoionization of atoms in a beam of natural abundant
calcium~\cite{mortensen04,herskind08}. The Coulomb crystals are
created through Doppler laser-cooling along the trap axis by two
counter-propagating beams at 397 nm (total power of $\sim 4$ mW and
beam diameter of $\sim$ 1 mm) and an 866 nm beam applied from the
side to prevent the ions from being shelved into the D$_{3/2}$ state
(see insert of Fig.~\ref{fig1}). The number of ions constituting the
crystals can be deduced by imaging the light (397 nm) scattered by
the ions during the cooling process onto a charge coupled device
(CCD) chip~\cite{herskind08}.

Normal modes of the ion Coulomb crystals can be excited by applying
identical AC electrical potentials at variable frequency to the four
end-pieces at each end of the trap. By this simple geometry modes
corresponding to the $(l,m=0)$ charged liquid modes can easily be
excited by having the voltages applied at each end either in-phase
($l$ even) or 180$^{\circ}$ out of phase ($l$ odd).

To monitor the coherent collective resonant response of the atomic
ions constituting the crystal, a 11.8 mm-long optical cavity with a
measured TEM$_{00}$ mode waist of 37 $\mu$m is
used~\cite{herskind08}. The optical finesse of the cavity is about
3,000 at the 866nm resonant wavelength of the 3d~$^2$D$_{3/2}$~-~4p~
$^2$P$_{1/2}$ transition in the $^{40}$Ca$^+$ ion addressed in the
experiments.

The plasma mode diagnostic involves a sequence of steps. First, a
Doppler cooling period of 5 $\mu$s is followed by a period of 12
$\mu$s, where the ions in the crystal are prepared in the $m_J=+3/2$
magnetic sub-state of the long-lived metastable D$_{3/2}$ level by
optical pumping (efficiency $\sim 97\%$). Next, a weak left-handed
circularly-polarized pulse (1.4 $\mu$s long) of 866 nm light is
coupled into the cavity to probe the collective response of the
ions. The mean intracavity photon number is less or about one at any
time. During this probing period, the photons reflected by the
cavity are measured by an avalanche photodiode with an overall
collection efficiency of 16\%. This sequence is repeated at a rate
of 50 kHz while the cavity length is scanned at a rate of 30 Hz. The
cavity lineshape is reconstructed by averaging a few hundred scans
(see \cite{herskind09} for more details).

With the ion Coulomb crystal at rest, the cavity probe linewidth is
given by~\cite{herskind09}:
\begin{equation}\label{eq:kappa_orig}
\kappa'=\kappa+g_N^2\frac{\gamma}{(\gamma^2+\Delta^2)},\end{equation}
where $\kappa$ is the cavity decay rate, $g_N$ the collective
coupling rate, $\gamma$ the optical dipole decay rate and $\Delta$
the detuning of the probe with respect to atomic resonance. If the
crystal is moving with a fixed velocity $v_z$ along the cavity axis
the probe linewidth would due to the Doppler shift be given
by
\begin{equation}\label{eq:kappa}
\kappa'=\kappa+g_N^2\frac{\gamma(\gamma^2+\Delta^2+(kv_z)^2)}{(\gamma^2+\Delta^2)^2+2(\gamma^2-\Delta^2)(kv_z)^2+(kv_z)^4}\end{equation}
where $k$ is the wavevector of the cavity field.

When exciting one of the normal modes, all the ions in the crystal
will generally not have the same velocity at a given instance (this
is only true for the center-of-mass mode, see
Eq.~(\ref{eq:deltaz})), and, furthermore, the velocity of the
individual ions will change during the mode-period. However, if the
timescale at which the velocity of the individual ions changes is
slow as compared to the effective cavity photon decay time, one can
model the expected probe lineshape by averaging the contributions
from the different parts of the crystal over a single
mode-oscillation period. From Eqs.~(\ref{eq:kappa_orig}) and
(\ref{eq:kappa}) it follows that any excitation of the ions' motion
will generally lead to a narrower linewidth due to the reduced
coupling strength.

Fig.~\ref{fig2}b shows the cavity probe linewidth for $\Delta$=0 as
a function of the modulation frequency applied to excite the
equivalent of the $(2,0)$ ``quadrupole" mode of a 1.2 mm-long
crystal with 4000 ions at a density $2.6\times 10^8$ cm$^{-3}$
(Fig.~\ref{fig2}a). A clearly reduced linewidth is observed around
142 kHz. The exact resonance value of
$\omega_{(2,0)}^{meas}/2\pi=142.1\pm 0.1$~kHz is in good agreement
with the resonance frequency $\omega_{(2,0)}^{model}/2\pi=142.2\pm
1.1$~kHz expected for the $(2,0)$ mode of a charged liquid crystal
for this aspect ratio and charge density.
\begin{figure}
  \includegraphics[width=8cm]{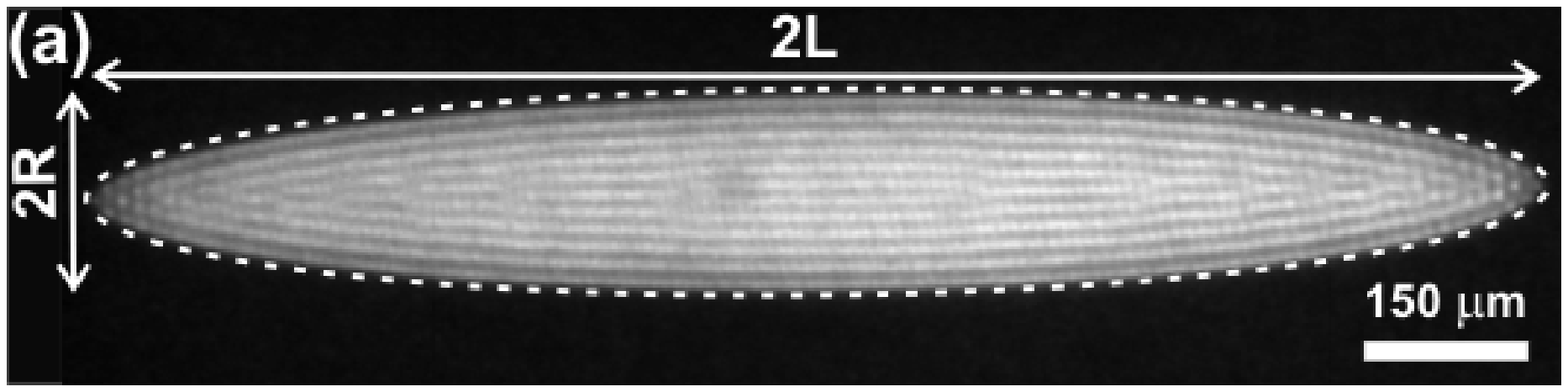}\\
  \includegraphics[width=7.9cm]{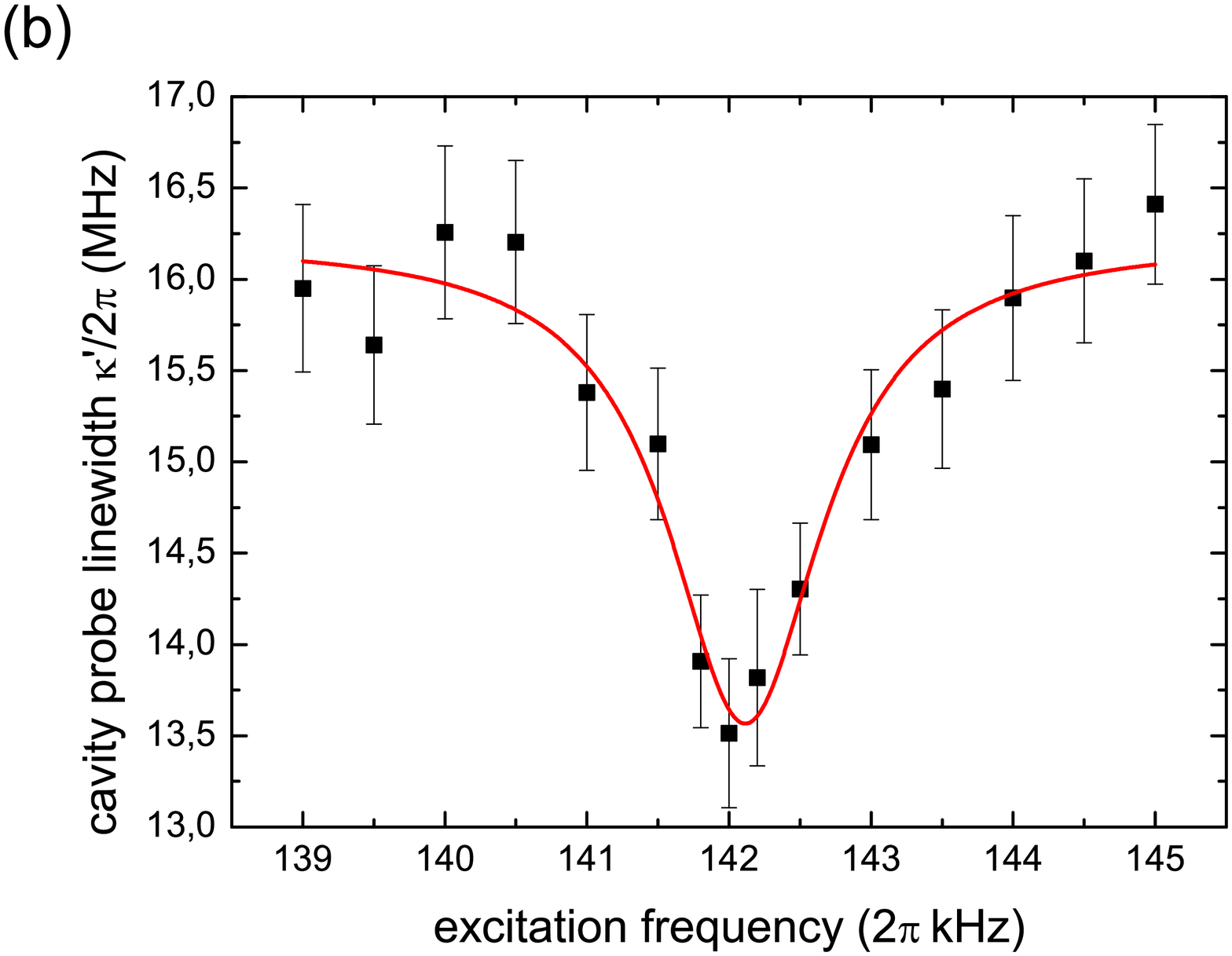}
  \caption{(a) Projection image of a 1.2 mm-long crystal with 4000 ions and density $2.6\times 10^8$ cm$^{-3}$.
  The DC and rf voltage amplitudes are 2.36 V and 205 V, respectively, and the aspect ratio $\alpha=R/L$ is 0.135. (b) Cavity probe linewidth as a function of the mode excitation frequency applied
  to drive the equivalent of the $(2,0)$ mode of this crystal. The probe is tuned to atomic resonance ($\Delta=0$) and the AC modulation voltage is 0.75 V. The red line is a
  Lorentzian fit.} \label{fig2}
\end{figure}
In order to test more generally how well the uniformly charged
liquid model describes unmagnetized ion plasmas confined in a linear
Paul trap, the resonance frequencies of the lowest lying normal
modes of ion crystals with various aspect ratios have been
determined by monitoring the linewidth of the cavity probe field
when tuned to the atomic resonance. The resulting mode resonance
frequencies are presented in Fig.~\ref{fig3} together with the
predicted values from the charged liquid model (see
Eq.~(\ref{eq:modefrequencies})). The measured values are consistent
with the theory to better than one percent for all experimental
data. This accuracy may appear quite surprising considering that
during all these $(l,0)$ mode measurements, the $(2,2)$ mode is
continuously off-resonantly excited by the linear rf quadrupole
field confining the plasma. The rf-induced modulation depth is
indeed up up to 20\% of the radial extension of the crystal, i.e.
comparable to that typically used for the axial excitation of the
$(l,0)$ modes. The inset of Fig.~\ref{fig3} clearly shows, however,
that, within the current experimental accuracy, the $(2,0)$ mode
frequency does not have any systematic dependence on the $(2,2)$
mode modulation depth $\xi_{(2,2)}^{rf}$, defined by the micromotion
amplitude of an ion at the radial position $(x_0,y_0)$
$(x(t),y(t))=(x_0(1+\xi_{(2,2)}^{rf}\cos \Omega t),
y_0(1-\xi_{(2,2)}^{rf}\cos \Omega t))$. This result is though
consistent with molecular dynamics simulations from which it has
been predicted that the radial rf field-driven micromotion in linear
Paul traps should have an extremely weak coupling into the axial
motion of the ions~\cite{schiffer00}.
\begin{figure}
  \includegraphics[width=8cm]{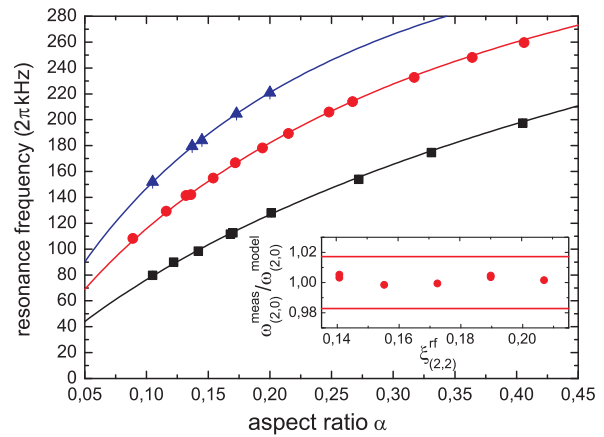}
  \caption{Resonance frequencies corresponding to the $(1,0)$ (squares), the $(2,0)$ (circles) and the $(3,0)$ (triangles) modes as a function of the aspect ratio $\alpha$,
  for a fixed plasma frequency $\omega_p/2\pi=536$ kHz ($U_{rf}=205 V$).
  The solid lines show the theoretical predictions of Eq.~(\ref{eq:modefrequencies}). The error bars are within the point size.
 The inset shows the $(2,0)$ mode resonance frequency as a function of the $(2,2)$ mode off-resonant modulation depth $\xi_{(2,2)}^{rf}$ (see text) for a fixed aspect ratio $\alpha=0.135$.
 The mode resonance frequency is normalized to that expected for a plasma without any excitation of the $(2,2)$
 mode and
 the red lines show the uncertainty in the expected resonance frequency due to the density calibration.
  The Coulomb crystals contain between 4000 and 12000 ions, with densities of $2.6-5.6\times 10^8$ cm$^{-3}$ and their temperature is in the 10 mK range.} \label{fig3}
\end{figure}

\begin{figure}
  \includegraphics[width=8cm]{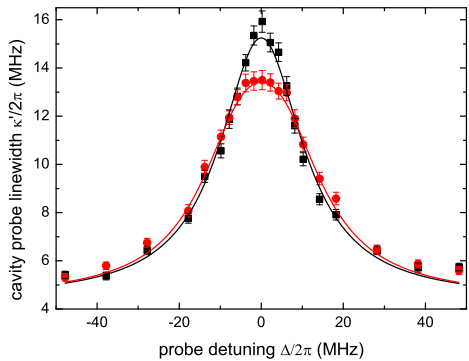}
  \caption{Cavity probe linewidth as a function of the probe
detuning, with (circles) or without (squares) modulation at the
$(1,0)$ mode resonance frequency, for a crystal similar to the one
presented in Fig.~\ref{fig2}. The solid lines are fits according to
the theoretical model described in the text.} \label{fig4}
\end{figure}

In Fig.~\ref{fig4}, the cavity probe linewidth with and without
exciting the $(1,0)$ mode at the resonance frequency
$\omega_{(1,0)}/2\pi=94$ kHz is presented as a function of the
detuning of the probe with respect to the atomic resonance. From the
Lorentzian line profile in absence of the mode excitation, one finds
$(\kappa,\gamma,g_N)/2\pi=(2.2\pm 0.1,12.3\pm 0.3,8.2\pm 0.1)$ MHz.
When the $(1,0)$ mode is excited, the absorption line is
substantially modified to a non-Lorenzian profile. A fit based on
the $(1,0)$ mode function (Eq.~(\ref{eq:kappa})) averaged over one
oscillation period results in a driven motion amplitude
$v_{z0}^{fit}=5.9\pm 0.7$ m/s
($v_{z}(t)=v_{z0}\cos(\omega_{(1,0)}t+\varphi)$). This value is in
good agreement with the value $v_{z0}^{fluo}=5.3\pm 0.6$ m/s deduced
from fluorescence images recorded in phase with the modulation
voltage. Here, a motional amplitude $z_0^{fluo}=9\pm 1 $ $\mu$m was
measured at the mode resonance frequency $\omega_{(1,0)}/2\pi=94$
kHz. Since equally good quantitative agreement between experiments
and the model was also found for the $(2,0)$ mode, this proves that
quantitative information about the ions' motion can reliably be
obtained from the ion-cavity coupling without the need for observing
directly the fluorescence signal.
In the future, higher order $(l,0)$ modes are envisioned to be
studied through mode excitations using spatial- and time-modulated
radiation pressure forces. For $^{40}$Ca$^+$ ions this can e.g. be
achieved through the combined application of a 866 nm repumper beam
with a spatially modulated intensity profile along the cavity axis
and a time-varying intensity of one of the 397 nm cooling
beams~\cite{drewsen04}. Eventually, for the high spatial modulation
of modes with large $l$, the liquid model should cease to apply.
Further applications could be measurements of ion Coulomb crystal
temperatures and heating rates~\cite{jensen05} and more detailed
investigations of the coupling between the various normal modes at
various temperatures and structural phases of the
plasma~\cite{dubin96,schiffer96}. Finally, we also believe that the
spectroscopic findings as well as the non-invasive character of the
method used will be important for e.g. observing radiation
pressure-induced optomechanical effects~\cite{murch08,brenneke08}
with ion Coulomb crystals.

We acknowledge financial support from the Carlsberg Foundation and
the Danish Natural Science Research Council through the ESF EuroQUAM
project CMMC.\\

\end{document}